%% file: size.tex
\pdfoutput=1

\documentclass[10pt,letter,twocolumn]{article}
\usepackage{f1000_styles}

\usepackage[super,comma,sort]{natbib}
\usepackage[colorlinks=true,urlcolor=blue]{hyperref}

\usepackage{amsmath}
\input size-def

\def\citeyear{\citep}
\def\autocite{\citep}
\def\textcite{\citet}

\begin{document}

\title{The invariances of power law size distributions}
\author[1]{Steven A.~Frank}
\affil[1]{Department of Ecology and Evolutionary Biology, University of California, Irvine, CA 92697--2525 USA, safrank@uci.edu}

\maketitle
\thispagestyle{fancy}

\parskip=4pt

\begin{abstract}

Size varies. Small things are typically more frequent than large things. The logarithm of frequency often declines linearly with the logarithm of size. That power law relation forms one of the common patterns of nature. Why does the complexity of nature reduce to such a simple pattern? Why do things as different as tree size and enzyme rate follow similarly simple patterns?  Here I analyze such patterns by their invariant properties. For example, a common pattern should not change when adding a constant value to all observations. That shift is essentially the renumbering of the points on a ruler without changing the metric information provided by the ruler. A ruler is shift invariant only when its scale is properly calibrated to the pattern being measured. Stretch invariance corresponds to the conservation of the total amount of something, such as the total biomass and consequently the average size. Rotational invariance corresponds to pattern that does not depend on the order in which underlying processes occur, for example, a scale that additively combines the component processes leading to observed values. I use tree size as an example to illustrate how the key invariances shape pattern. A simple interpretation of common pattern follows. That simple interpretation connects the normal distribution to a wide variety of other common patterns through the transformations of scale set by the fundamental invariances. 

\end{abstract}


\vskip0.5in
\noterule
Preprint of published version: Frank, S. A. 2016. The invariances of power law size distributions. F1000Research 5:2074,  \href{http://dx.doi.org/10.12688/f1000research.9452.2}{doi:10.12688/f1000research.9452.2}. Published under a Creative Commons \href{https://creativecommons.org/licenses/by/4.0/}{CC BY 4.0} license.
\smallskip
\noterule

\clearpage

\section*{Introduction}

The size of trees follows a simple pattern. Small trees are more frequent than large trees. The logarithm of frequency declines linearly with the logarithm of size\autocite{farrior16dominance}. Log-log linearity defines a power law pattern. Power laws are among the most common patterns in nature\autocite{mandelbrot83the-fractal}.

Power laws arise by aggregation over a multiplicative process, such as growth. Many processes in nature apply a recursive repetition of a simple multiplicative transformation, with some randomness\autocite{mandelbrot83the-fractal}. 

Aggregation over a random multiplicative process often erases all information except the average logarithm of the multiplications \autocite{frank09the-common,frank14how-to-read}. That average determines the slope of the power law line. In the case of tree size, we must also account for the fact that trees cannot grow to the sky. The upper bound on growth causes the frequencies of the largest trees to drop below the power law line.

That simple view of aggregation and the regularity of power laws contrasts with an alternative view. By the alternative view, the great regularity of a power law pattern suggests that there must be a very specific and particular underlying generative process. If the pattern of tree size is so regular, then some specific process of trees must have created that regularity. 

To support the simple view of aggregation and regularity, I show that a normal distribution contains the same information as a power law size distribution. The distributions differ only in the scaling used to measure the distance of random variations in size from the most common size\autocite{frank16common}.

The normal distribution calls to mind the great regularity in pattern that arises solely from the aggregation of an underlying stochastic process. Stochasticity and aggregation alone are sufficient to explain the regularity\autocite{jaynes03probability}. There is no need to invoke a detailed generative process specific to trees. Given the observed power law of sizes, maybe all we can reasonably say is that growth is a stochastic multiplicative process and that trees do not grow to the sky.

The trees provide an example of deeper principles about pattern and process in biology. What exactly are those principles? How can we use those principles to gain insight into biological problems?

To start on those questions, the next section presents an example of tree size data. Those data follow a power law with an upper bound on size. I show that those data also match almost exactly to a normal distribution when scaled with respect to a natural metric of growth. 

The normal distribution and the power law pattern express the same underlying relation between pattern and process. That underlying relation arises from a few simple invariance principles. I introduce those invariance principles and how those principles shape the common patterns of nature\autocite{frank16common}.

\begin{figure*}
\centering
\includegraphics[width=1.0\hsize]{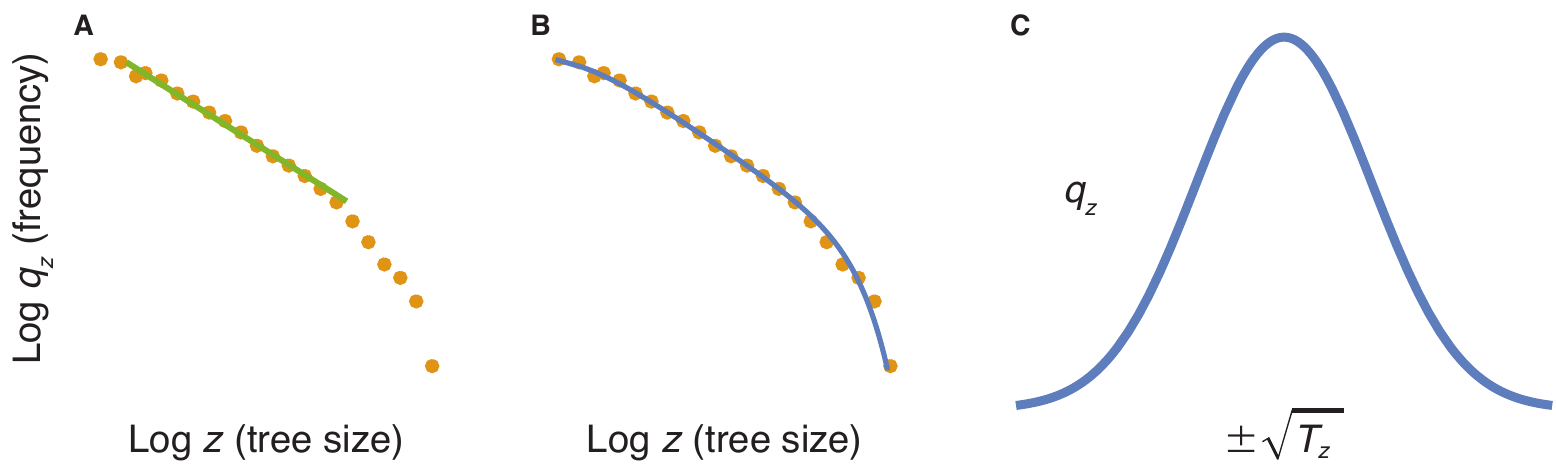}
\caption{(A) Tree size, $z=d^2$, in which the squared diameter, $d^2$, is proportional to the cross sectional area of the stem, and $d$ ranges over approximately $11$--$2800$mm. The green line shows great regularity of pattern as a power law over the range that covers almost all probability. The largest trees, beyond the green power law line, comprise only a small fraction of all trees, because of the logarithmic scaling of frequency. (B) The blue line is $\log q_z = \log k-\Gl \Tz$, with $\Tz= \log(1+az)+\Gg z$, and parameters $\Gl=1.06$, $a=0.004$, and $\Gg=7\times10^{-7}$, with $\log k$ shifting curve height and total probability. (C) The fitted blue line in panel B is a classic normal distribution with variance $1/2\Gl$ when plotted as $q_z \propto e^{-\Gl \Tz}$ versus $\pm\sqrt{\Tz}$, with respect to $z$ as a positive parameter. In this plot, the metric is shifted so that the most common type associates with a value $\Tz=0$. Data approximated from Figure 4 in \textcite{farrior16dominance}}
\label{fig:trees}
\end{figure*}

\section*{Tree size}

\Figure{trees}A shows the distribution of tree size in a tropical forest\autocite{farrior16dominance}. Most of the trees lie along the green power law. The largest trees, beyond the line, comprise only a small fraction of all trees, because of the logarithmic scaling of frequency.

The blue curve in \Fig{trees}B closely fits the observed pattern. That curve expresses the natural metric for variation in tree size, $z$, as 
\begin{equation}\label{eq:Tz}
  \Tz = \log(1+a z) + \Gg z.
\end{equation}
This metric relates size to a logarithmic term for multiplicative growth plus a linear term for an upper bound on size. There is no additional information in the fitted curve beyond this natural metric.

The normal distribution in \Fig{trees}C expresses exactly the same information about the distribution of tree sizes as the fitted curve in \Fig{trees}B. The normal distribution follows from the expression of size variation in terms of the natural metric, $\Tz$. I derive these conclusions in the following sections.

\section*{Natural metrics}

The pattern of tree size can be understood by considering $\Tz$ as a natural metric for size. A natural metric expresses a shift and stretch invariant scale for an observed probability pattern \autocite{frank16common}. Shift, by adding a constant to a natural metric, does not change observed pattern. Stretch, by multiplying the metric by a constant, does not change pattern.

Ideally, a natural metric also expresses the relation between underlying process and observed pattern. However, we can be right about the proper natural description of observed pattern but wrong about its underlying cause. It is important to distinguish description from causal interpretation. 

The next section describes the natural metric for tree size with respect to the fundamental invariances of shift and stretch. I discuss the panels of \Fig{trees} as simple expressions of the natural metric.  The following sections consider how to interpret natural metrics, the description of observed pattern, and the analysis of underlying process. The presentation here extends the underlying abstract theory to the interpretation and intuitive understanding of empirical pattern. Technical details can be found in the cited articles.

\section*{The metric of tree size: affine invariance}

The data\autocite{farrior16dominance} in \Fig{trees} arose from measurements of trunk diameter, $d$. I sought a natural metric based on $d$ that describes the data in a shift and stretch invariant manner\autocite{frank16common}. 

How does one find a shift and stretch invariant natural metric that matches an observed pattern? In practice, one uses the extensive underlying theory and prior experience in what often works\autocite{frank09the-common,frank10measurement,frank11a-simple,frank14how-to-read}. I achieved an excellent fit to the observed tree size data in \Fig{trees}B based on the metric, $\Tz$, in \Eq{Tz}. I summarize the steps by which I arrived at that metric.

The data form a probability distribution. Probability patterns have a generic form. Measurements, $z$, relate to the associated probability, $\qz$. The natural metric, $\Tz$, transforms measurements such that the probability pattern has the exponential form
\begin{equation}\label{eq:expdistn}
  \qz = ke^{-\Gl\Tz},
\end{equation}
in which $\Gl$ adjusts the stretch of $\Tz$, and $k$ adjusts the total probability to be one. 

Probability patterns in the exponential form are shift and stretch invariant with respect to the metric, $\Tz$. In particular, the affine transformation of shift and stretch, $\Tz\mapsto \Ga+\Gb\Tz$, is exactly compensated by adjustments of $k$ and $\Gl$, leaving the probability pattern invariant.

Intuitively, we can think of affine invariance as defining a ruler that is linear in the metric, $\Tz$. In a linear ruler, it does not matter where we put the zero point. The information in measurement depends only on the distance from where we set zero to where the observation falls along the ruler. That independence of the starting point is shift invariance. 

Similarly, if we uniformly stretch or shrink the ruler, we still get the same information about the relative values of different measurements. All we have to do is multiply all measurements by a single number to recover exactly the same distances along the original ruler. The metric $\Tz$ provides information that is stretch invariant.

To fit the data of \Fig{trees}A, we have to find the matching affine invariant metric, $\Tz$, for probability expressed in the exponential form of \Eq{expdistn}. 

\section*{The metric of tree size: scale}

Most natural metrics are simple combinations of linear, logarithmic, and exponential scaling\autocite{frank11a-simple,frank14how-to-read}. For example, in the metric $\Tz= \log z + \Gg z$, the logarithmic term dominates when $z$ is small, and the linear term dominates when $z$ is large. The metric scales in a log-linear way. Change in scale with magnitude often occurs in natural metrics. 

Roughly speaking, the linear, logarithmic, and exponential scales correspond to addition, multiplication, and exponentiation. Those arithmetic operations are the three primary ways by which quantities combine. One can think of numbers combining additively, multiplicatively or exponentially at different magnitudes, depending on the way in which process changes with magnitude. 

Small trees tend to grow multiplicatively, and large trees tend to scale linearly as they approach an upper size limit. \textcite{farrior16dominance} used logarithmic scaling at small magnitudes and linear scaling at large magnitudes. However, they did not express a metric that smoothly changed the proportion of the two scalings with magnitude. Instead, they switched from log to linear scaling at some transition point. 

The observed data fit roughly to a pure log-linear metric, $\Tz= \log z + \Gg z$, with $z=d$ as tree diameter. I obtained a better fit by modifying this metric in two ways to obtain the expression in \Eq{Tz}. 

First, I used the square of the diameter, $z=d^2$, which is proportional to the cross sectional area of the trunk at the point of measurement. Various intuitive reasons favor area rather than diameter as a measure of size and growth. However, I ultimately chose area because it fit the data. 

Second, I replaced $\log z$ by $\log(1+a z)$. On a pure log scale, $\log z$ explodes to negative infinity as $z$ approaches zero. In application to positive data, such as size, it almost always makes sense to use $\log(1+a z)$. This expression becomes smaller in magnitude as $z$ declines. The parameter $a$ scales the rate of change with respect to the point of origin. 

Size distributions often follow the metric, $\Tz=\log(1+az)+\Gg z$. Of course, not all distributions follow that pattern. But one can use it as a default. When observations depart from this default, the particular differences can be instructive.

\section*{Interpretation of natural metrics}

The natural metric of a probability pattern transforms observed values on the scale $z$ into probability values on the scale $\Tz$. Through the natural metric, the particular pattern on the observed scale, $z$, becomes a universal probability pattern in the natural metric, $\Tz$. 

One can understand the intuitive basis of natural metrics by considering the properties of the universal probability scale. Probability patterns are often discussed with words such as \emph{information} or \emph{entropy}\autocite{cover91elements}. Those words have various technical and sometimes conflicting definitions. But all approaches share essential intuitive concepts. 

\emph{Surprise} expresses the intuition\autocite{tribus61thermostatics}. Rare events are more surprising than common events. Suppose a particular size, $z$, occurs in one percent of the population, and another size, $z'$, occurs in two percent of the population. We will be more surprised to see $z$ than $z'$. How much more surprised? 

Surprise is relative. We should be equally surprised by comparing probabilities of $0.01$ versus $0.02$ and $0.0001$ versus $0.0002$. Each contrast compares one event against another that is twice as common. 

What is a natural metric of probability that captures these intuitive notions of surprise? For probability, $\qz$, the surprise is defined as
\begin{equation}\label{eq:surprise}
  \sz = -\log\qz.
\end{equation}
We compare events $z$ and $z'$ by taking the difference
\begin{equation*}
  \sz - \surp_{z'} = \log q_{z'} - \log\qz = \log\frac{q_{z'}}{\qz}.
\end{equation*}
This natural metric, $\sz$, leads to affine invariant comparisons of surprise values. In the affine transformation, $\surp\mapsto\Ga+\Gb\surp$, the shift $\Ga$ cancels in the difference $\sz - \surp_{z'}$. The stretch $\Gb$ causes a constant change in length independently of location, so the metric retains the same information at all magnitudes of the scale. 

The relation between the universal metric of probability, $\sz$, and the natural metric for a particular observed scale, $\Tz$, follows from the exponential form for probability\autocite{frank16common} in \Eq{expdistn}. From that exponential form, we can write $\sz = \Gl\Tz - \log k$. Because $\sz$ is shift invariant, we can ignore the constant $\log k$ term, yielding
\begin{equation*}
  \sz = \Gl\Tz.
\end{equation*}
The natural metric, $\Tz$, transforms an observed scale, $z$, into the universal metric of probability patterns, $\sz$. The fitted curve in \Fig{trees}B is a plot of $\sz = \Gl\Tz$ versus $\log z$.

To interpret a scale, it is useful to think about what happens along each increment of the scale. Define $\dd\sz$ and $\dd\Tz$ as small increments along the scales at the point associated with $z$. Then
\begin{equation*}
  \dd\sz = \Gl\dd\Tz,
\end{equation*}
which means that the scales $\sz$ and $\Tz$ change in the same way at all magnitudes of $z$, with $\Gl$ as the constant of proportionality in the translation from one scale to the other. 

How do small increments in the natural metric, $\dd\Tz$, relate to increments in the observed values, $\dd z$? If we assume that $\Tz$ increases with $z$, and define $\Tz'=\dd\Tz/\dd z$ as the derivative (slope) of $\Tz$ with respect to $z$, then 
\begin{equation*}
  \dd\sz = \Gl\Tz'\dd z.
\end{equation*}
Here $\Tz'$ transforms increments along the observable scale, $\dd z$, into increments along the universal scale of probability pattern, $\dd\sz$. All of the information that relates observation to probability pattern is summarized by the natural metric, $\Tz$. 

\section*{Generative process: generic vs particular}

What underlying generative process leads to an observed pattern? We must separate two aspects. Generic aspects arise from general properties of aggregation, measurement and scale that apply to all problems. Particular aspects arise from the special attributes of each problem. 

Confusing generic and particular aspects leads to the greatest misunderstandings of pattern and process\autocite{frank09the-common,frank14how-to-read}. For example, the observed pattern in \Fig{trees} perfectly expresses generic properties.  Aggregation leads to the normal distribution by the central limit theorem (\Fig{trees}C). The natural metric of size, $\Tz$, relates the normal distribution to power law and exponential scaling in \Fig{trees}A,B, when probability is plotted with respect the logarithm of the observed values, $z$. 

In the tree size data, simple generic properties account for all of the observed pattern. I do not mean that there is nothing particular about trees or that we cannot study how ecological processes influence tree size. I mean that we must not confuse the generic for the particular in our strategy of inference \autocite{frank09the-common,jaynes03probability,harte11maximum}.  

This article focuses on generic aspects of pattern. The following sections discuss those generic aspects in more detail.

\section*{The normal distribution and generic pattern}

One often observes great regularity in probability patterns. Tree size follows a power law with an upper bound. Other measurements, such as height, weight, and enzymatic rate, also express regularity, but with different patterns. 

A single underlying quantity captures the generic regularity in seemingly different patterns. That underlying quantity is the average distance of observations from the most common type\autocite{jaynes03probability}. The key is to get the correct measure of distance, which is the natural metric. 

The normal distribution is a pure expression of the generic regularity in probability patterns. In the normal distribution, the variance is the average distance of fluctuations from the mean.

In the normal distribution, the natural metric is the squared deviation from the mean, $\Tz=z^2$. Here, $z$ is the observed deviation from the mean, and $\Tz$ is the natural metric for distance. The normal distribution follows from the standard expression of probability patterns in \Eq{expdistn}, repeated here with $v=k$, as
\begin{equation}\label{eq:expnormal}
  \qz=ve^{-\Gl\Tz}.
\end{equation}
The average of the squared deviations, $\Tz=z^2$, is the average distance of fluctuations from the most common type, which is the definition of the variance, $\Gs^2$. We can express the parameters in terms of the variance
\begin{equation}\label{eq:normalparam}
  \Gl = \frac{1}{2\Gs^2} \mskip80mu v = \sqrt{\frac{\Gl}{\pi}}=\frac{1}{\sqrt{2\pi\Gs^2}},
\end{equation}
from which we derive the commonly written form for the normal distribution as
\begin{equation}\label{eq:normal}
  \qz=\frac{1}{\sqrt{2\pi\Gs^2}}\,e^{-z^2/2\Gs^2}.
\end{equation}
The normal distribution is universally known but rarely understood. Interpreting the powerful generic aspect of probability patterns often reduces to correctly reading this equation. 

The standard expression for the normal distribution in \Eq{normal} seems obscure. By understanding that \Eq{expnormal} expresses the same information in a much more general and broadly applicable way, we learn to read the simple generic aspect of common pattern. The key arises from the relation between the natural metric, $\Tz$, and the measurement scale, $z$, used to express the pattern.

\section*{Metrics of probability and measurement}

This section discusses key aspects of the natural metric transformations, $\Tz$, of the underlying measurements, $z$. The understanding of probability pattern arises from these key aspects of the natural metric.

Suppose that two observers measure the same pattern. One uses a ruler that follows the scale, $z$. Another has a logarithmic ruler that returns logarithmic values, $\log z$, for the same underlying values. The two observers do not know that they are using different scales.

When the two observers plot their data, each will see a different probability pattern. The plot of $\qz$ versus $z$ differs from the plot of $\qz$ versus $\log z$. 

Similarly, two observers may see different patterns of human size if they measure different things. Suppose one observer measures femur length, the other measures cross sectional area of the chest. The probability patterns of femur and chest size differ. But the different patterns reflect the same information about the underlying size variation in the population. 

What is the best way to find the relation between different observed values and the common underlying information about variation? Often, the natural metric for each observed scale provides the universally comparable scale for probability pattern. That universally comparable scale can be used to express variation as a normal distribution.

When an observed probability pattern matches the normal distribution, then the variance summarizes all of the information in the pattern\autocite{jaynes03probability}. We can write the variance, $\Gs^2$, which is the average of the squared distance for fluctuations from the mean, as
\begin{equation*}
  \Gs^2 = \angb{z^2}_z
\end{equation*}
in which the angle brackets denote the average value of $z^2$, and the subscript $z$ means that the average is taken with respect to the underlying scale, $z$. 

The great generality of the normal distribution arises from a broader concept of the average distance of fluctuations from a central location
\begin{equation}\label{eq:var}
  \Gs^2 = \angb{z^2}_z \mskip20mu\longrightarrow\mskip20mu \tilde{\Gs}^2 = \angb{\T\mmskip}_{\sqrt{\T}}.
\end{equation}
The left shows the standard definition of the variance as the average squared distance from a central location. The right generalizes that notion of average squared distance by using the average of the natural metric, $\Tz$, in which the average is taken with respect to the square root of the natural metric, $\sqrt{\Tz}$. Here, $\Tz$ is shifted so that the most common type associates with $\Tz=0$, and the metric expresses fluctuations from the most common type \autocite{frank16common}.

On the left, we average $z^2$ with respect to $z$. On the right, we average $\Tz$ with respect to $\sqrt{\Tz}$. The general form on the right-hand side includes the left-hand side as the special case of $\Tz=z^2$. 

The key conclusion is that common probability patterns expressed in their natural metric
\begin{equation*}
  \qz=ve^{-\Gl\Tz}
\end{equation*}
are normal distributions when plotting $\qz$ versus $\pm\sqrt{\Tz}$. 

The following sections present examples. Later sections show why the square root is a natural measurement scale for common probability patterns.

\section*{Natural metrics and generic forms}

The tree size data match almost perfectly to the generic normal distribution (\Fig{trees}C). I discuss that match in terms of universal properties of the normal distribution, given in the prior sections.

Tree size variation follows a simple log-linear natural metric, $\Tz$. That metric and its associated probability pattern 
\begin{equation*}
  \qz=ke^{-\Gl\Tz}
\end{equation*}
closely fit the data. \Figure{trees}B shows the fit when plotting $\log\qz$ versus $\log z$.  \Figure{trees}C shows that the same observed variation closely fits a normal distribution when plotting $\qz$ versus $\pm\sqrt{\Tz}$. 

The generalized variance is the average squared fluctuation of tree size from the most common type, when squared fluctuations are expressed by the natural metric, and fluctuations are measured by the square root of the natural metric. By the generalized notion of the variance in \Eq{var}, all of the information in the observed distribution of tree size is contained in the average distance of fluctuations, measured in the natural metric.

The transformation of data into a normal distribution is sometimes considered a trivial step in the statistical analysis of significance levels. Here, in contrast, the natural metric and the associated expression in normal form provide an essential step in the general understanding of pattern and process. 

Later sections discuss why the normal distribution arises as the simple expression of pattern in relation to natural metrics. Before turning to those concepts, I present another example.

\begin{figure*}
\centering
\includegraphics[width=0.8\hsize]{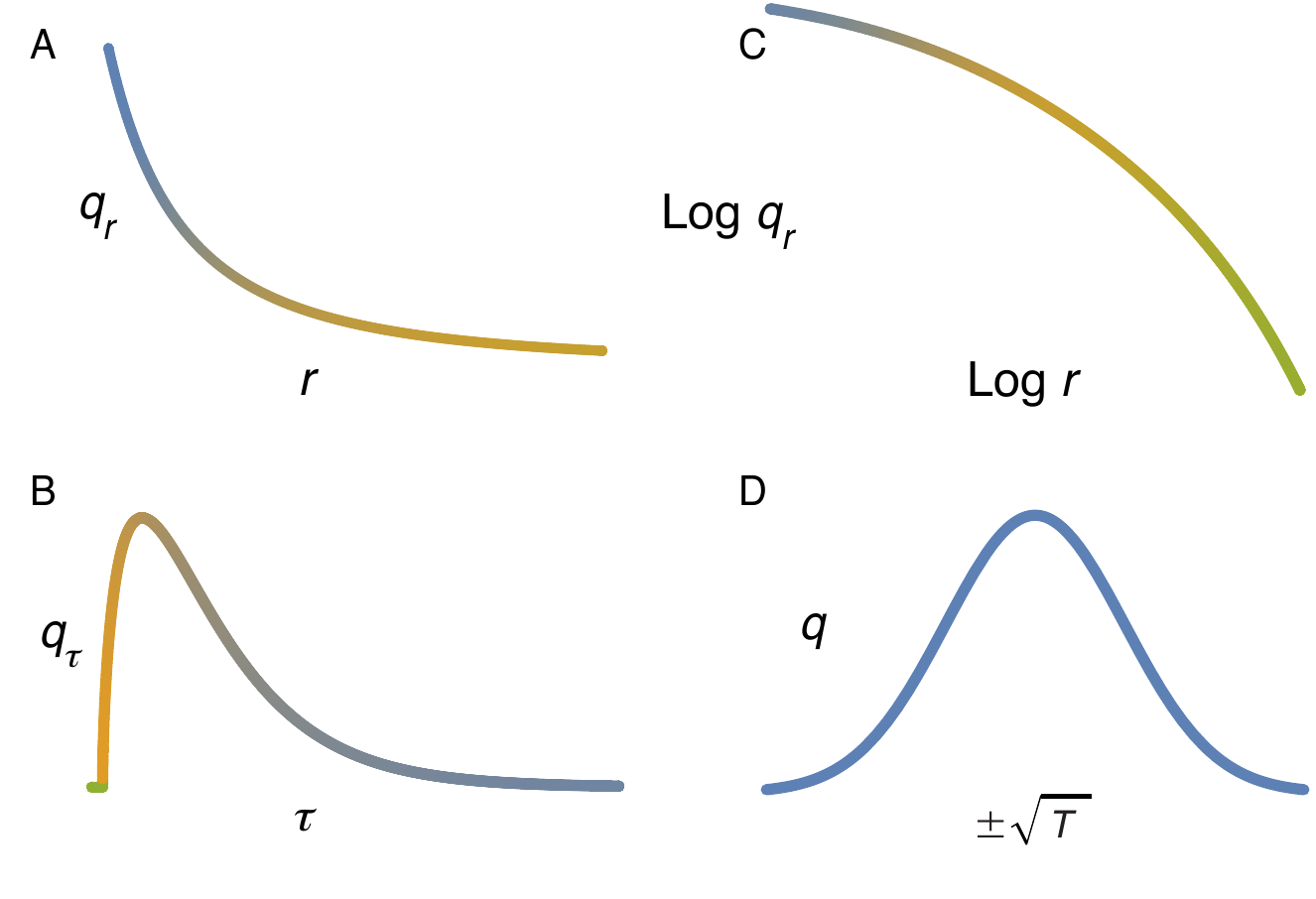}
\caption{A pair of common natural metrics related by dimensional inversion, with generic expression by the normal distribution. (A) The probability distribution based on the natural metric in \Eq{Tz}, with $\Tr = \log(1+ar) + \Gg r$. This plot uses a linear abscissa, compared with the logarithmic abscissa of \Fig{trees}A. The curve approximately fits the enzymatic rate data in Figure 2B of \textcite{iversen14ras-activation-short}, in which $r$ has units $S/t$ measured as number of molecules per unit time (seconds). Here, $r$ varies between $0$ and $8$. The approximately fitted parameters are $a=0.5$, $\Gg=0.05$, and $\Gl=1.6$. (B) The Laplace transform of the upper panel yields a shifted gamma probability distribution that expresses the identical information with a natural metric $\Tt = (1/\Gl-1)\log(\Gt-\Gg\Gl)+\Gt/\Gl a$. The inverted measure $\Gt$ has units $t/S$ as time per molecule, varying in the plot between $\Gg\Gl$ and $4$. (C) The same probability distribution as in panel A, on a double log scale over the range of $r$ values $0.6$ to $50$. (D) Both the original distribution in A and the Laplace inverted distribution in B are normal distributions when expressed in relation to the square root of their respective natural metrics, with generalized variance $\tilde{\Gs}^2$ in \Eq{var}. }
\label{fig:iversen}
\end{figure*}

\begin{figure}
\centering
\includegraphics[width=0.75\hsize]{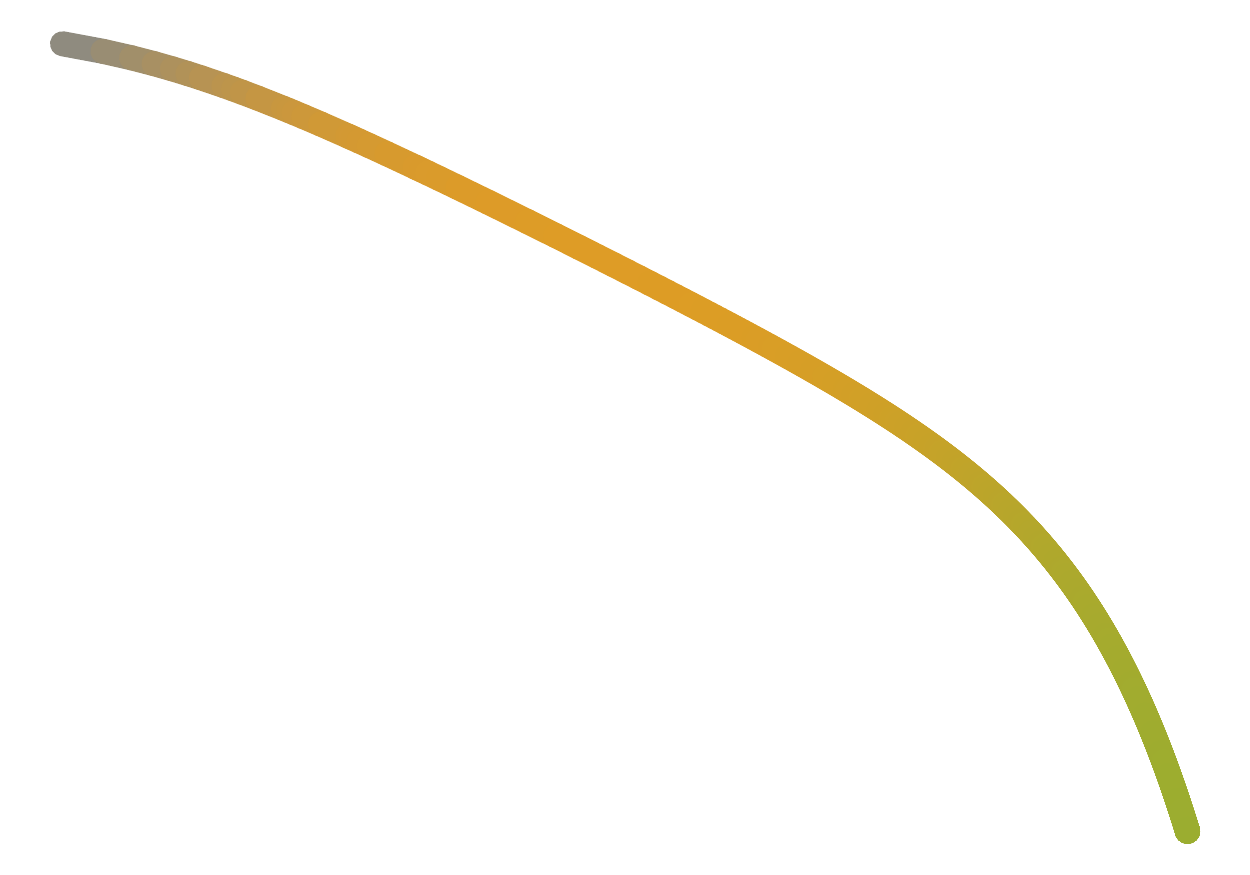}
\caption{The fitted probability distribution for the tree size data in \Fig{trees}B. This distribution has the same natural metric as in \Fig{iversen}C, but with different parameters. The curve is colored to show the change in the scaling of the natural metric with increasing magnitude as linear (blue), logarithmic (gold), and linear (green).}
\label{fig:treecolor}
\end{figure}

\section*{Dimensional inversion and metric pairs}

Natural metrics sometimes come in pairs\autocite{frank10measurement,frank14how-to-read}. For example, rates and frequencies follow dual metrics. Rates have dimensional units $S/t$, in which $S$ is a generic size or number unit, and $t$ is a time unit. A growth rate for trees may be given in terms of the change in size per year. A chemical reaction rate may be given as the number of molecules produced per unit time. 

The inverse of a rate has units $t/S$. That inverse expresses the time to grow larger or smaller by a particular size unit, or the time to produce a particular number of molecules. 

This section illustrates the common dual metrics for rates and times. The dual metrics yield different probability patterns that contain exactly the same underlying information. Each metric takes on the same common normal distribution form when stochastic fluctuations are measured by the metric relative to its square root. 

To illustrate the dual metrics, I use the measured rates of chemical reactions for individual enzyme molecules given by \textcite{iversen14ras-activation-short}. The measurements produce a probability pattern for the distribution of reaction rates. The measurements are not sufficiently precise to determine exactly which natural metric fits the data. 

I made an approximate fit to the data by using the natural metric in \Eq{Tz}, which I previously used to fit tree size. My only purpose here is to illustrate typical aspects of rate and frequency patterns, rather than to over-analyze the limited data available in this particular study. 

\Fig{iversen}A shows the fitted distribution of reaction rates. The rates are in molecules per second, $r$, with units $S/t$. The colors in the curve express the change in the scaling relations of the natural metric as magnitude increases. The natural metric from \Eq{Tz}, repeated here with $r=z$, is
\begin{equation*}
    \Tr = \log(1+a r) + \Gg r.
\end{equation*}
When $r$ is small, linear scaling of $\Tr$ dominates, as shown by the blue coloring. As $r$ increases, logarithmic scaling dominates, as shown by the gold coloring. \Fig{iversen}C, covering a greater range of $r$ values, shows that further increase in $r$ leads to linear dominance of scale, as shown by the green color. The upper linearity expresses the bound on size or number. Trees do not grow to the sky. Reaction rates do not become infinitely fast. \Fig{treecolor} shows the tree size data colored by the linear-log-linear transitions. 

The probability pattern for rates, $S/t$, has a natural dual pattern expressed by inverted units for time, $t/S$. We can invert units by the Laplace transform\autocite{frank10measurement,frank14how-to-read}. The inversion leads to an altered probability pattern based on the natural metric 
\begin{equation*}
    \Gl\Tt = \Ga\log(\Gt-d)+\Gt/a,
\end{equation*}
with $\Ga=1-\Gl$ and $d=\Gg\Gl$. The parameters match the paired metric, $\Tr$. The common value of $\Gl$ shared by the paired distributions arises from the full expression for probability patterns in \Eq{expdistn}. The probability pattern for time, arising from $\Tt$, is a gamma distribution shifted by $d$. 

The time per molecules pattern in \Fig{iversen}B matches the dual enzyme rate pattern of molecules per time in \Fig{iversen}A. The dual distributions express the identical information. 

Dimensional inversion associates the various linear-log-linear scales between the two forms\autocite{frank10measurement,frank14how-to-read}. The linear, blue component at small magnitude in the upper panel matches the long blue tail at large magnitude in the lower panel. Put another way, slow rates, $r$, correspond to long waiting times, $\Gt$. 

In the top, the gold logarithmic component for high rates matches the lower gold component for short waiting times. For very high rates, $r$, we have to look at \Fig{iversen}C.  The upper green linear tail corresponds to the rapid decline in the probability of observing extremely high rates, associated with the natural upper bound on rates. The green upper bound on rates matches the green lower limit on times in \Fig{iversen}B. If extremely rapid rates of reaction, $r$, are very rare, then no reactions will produce molecules in very short time periods, $\Gt$. That limitation produces the green shift at small times in \Fig{iversen}B.

The dual natural metrics of rate, $\Tr$, and time, $\Tt$, correspond to similar expressions of the normal distribution\autocite{frank16common} in \Fig{iversen}D. In general, different probability patterns expressed in different metrics, $\T$, become normal distributions when fluctuations from the most common value are measured by $\pm\sqrt{T}$.

\begin{figure*}
\centering
\includegraphics[width=0.8\hsize]{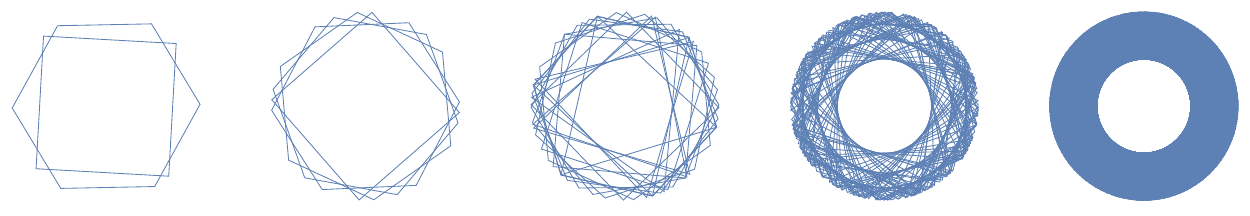}\vskip30pt
\includegraphics[width=0.8\hsize]{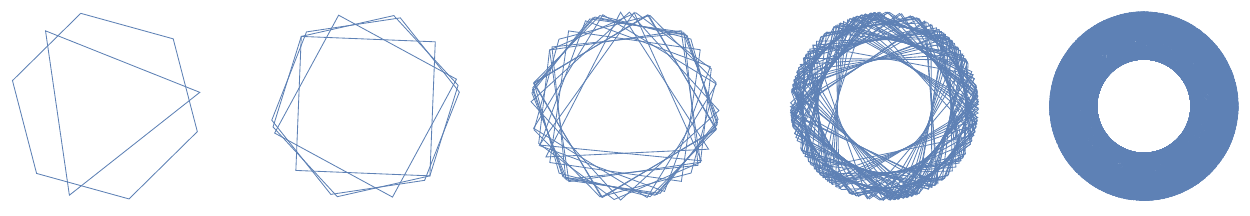}\vskip30pt
\includegraphics[width=0.8\hsize]{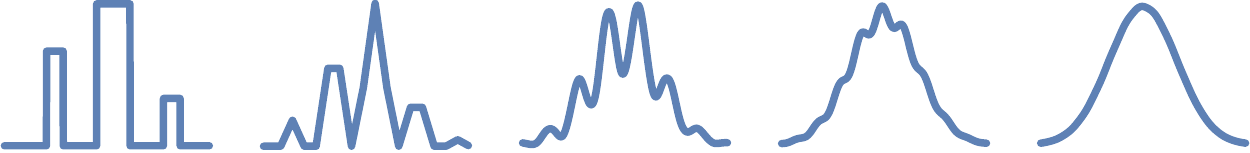}\vskip30pt
\includegraphics[width=0.8\hsize]{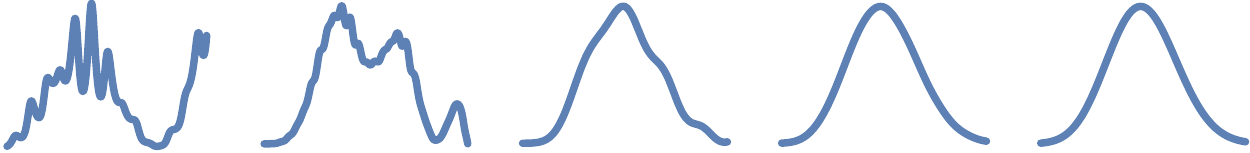}
\caption{Aggregation and asymptotic invariance. The top shows polygons randomly rotated about their center. Aggregation leads asymptotically to loss of all information about rotational orientation. A circle purely expresses that rotational invariance. The bottom shows the aggregate summing of observations from arbitrary probability distributions. Aggregates combine to produce normal distributions, purely expressing the loss of all information except the average distance (variance) from the most common observation. The normal distribution is invariant to the order in which observations are combined. Order invariance is similar to rotational invariance (\Fig{swiss}). Thus, the asymptotic circle and the asymptotic normal distribution express similar aspects of information loss and invariance.}
\label{fig:aggr}
\end{figure*}

\section*{Aggregation and asymptotic invariance}

Why do tree sizes and enzyme rates match a simple natural metric? Why do a few simple natural metrics match most of the commonly observed patterns? Part of the answer arises from the way in which aggregation leads to simple invariant pattern.

The top rows of \Fig{aggr} illustrate aggregation and invariance. Each row begins on the left with two regular polygons, randomly rotated about their center. Columns to the right add more randomly rotated components. As the random rotations aggregate, the shape converges asymptotically to an invariant circular form. 

Random rotation causes loss of information about the angle of orientation. In the aggregate, the asymptotic form is rotationally invariant. In other words, the circular shape remains invariant no matter how it is rotated. A circle expresses pure rotational invariance.  

The bottom two rows illustrate aggregation and the invariant pattern of the normal distribution. Each row begins on the left with a probability distribution. For each distribution, the horizontal axis represents observable values, and the vertical axis represents the relative probability of each observed value. I chose the shapes of the distributions to be highly irregular and to differ from each other. 

The second column is the probability distribution for the sum of two randomly chosen values from the distribution in the left column. The third, fourth, and fifth columns are, respectively, the sum of four, eight, and 16 randomly chosen values. The greater the aggregation of randomly chosen values, the more perfectly the pattern matches a normal distribution. Adding randomly chosen values often causes an aggregate sum to converge asymptotically to the invariant normal form.

\section*{Natural metrics and a universal scale}

The invariant normal form expresses a universal scale. That universal scale clarifies the concept of natural metrics. To understand the universal scale, we begin with the fact that the same pattern can be described in different ways. 

Consider enzyme catalysis. Fluctuations can be measured as the rate of molecules produced per unit time. Alternatively, fluctuations can be measured as the interval of time per molecule produced. \Fig{iversen}A, B show the dual expression of the same underlying information. 

The dual measurement scales each have their own natural metric. A natural metric transforms a particular measurement scale into a universal scale that expresses the common underlying information. A metric is \emph{natural} in the sense that it connects a particular scale of observation to a common universal scale. 

The normal distribution purely expresses the universal scale. Suppose we begin with different scales of measurement, such as the rate of molecules produced per unit time and the interval of time per molecule produced. Each scale has its own distinctive pattern of random fluctuations, as in \Fig{iversen}A, B. When we transform each scale to its natural, universal metric, $\Tz$, the pattern of random fluctuations follows the normal distribution (\Fig{iversen}D).

A normal distribution expresses information only about the average distance of fluctuations from the most commonly observed value. If we measure distance for different underlying measurements in their natural metrics, then that distance is the universal form of variance in \Eq{var} as
\begin{equation*}
  \tilde{\Gs}^2 = \angb{\T\mmskip}_{\sqrt{\T}}.
\end{equation*}
The generalized variance expresses the average deviation of the natural metric relative to the square root of the natural metric. 

Why is the relation between a natural metric and its square root the universal measure of scale and also the expression of the normal distribution? The answer concerns how rotation and aggregation lose information and leave an invariant pattern (\Fig{aggr}). 

The next section discusses rotational invariance and its relation to the universal scaling of the normal distribution. The following sections return to tree size and other commonly observed size distributions. The concepts of rotational invariance and the normal distribution clarify why the natural metric for tree size, given in \Eq{Tz} as $\Tz=\log(1+az)+\Gg z$, is a common natural metric for size patterns. 

\begin{figure}
\centering
\vskip5pt
\includegraphics[width=0.3\hsize]{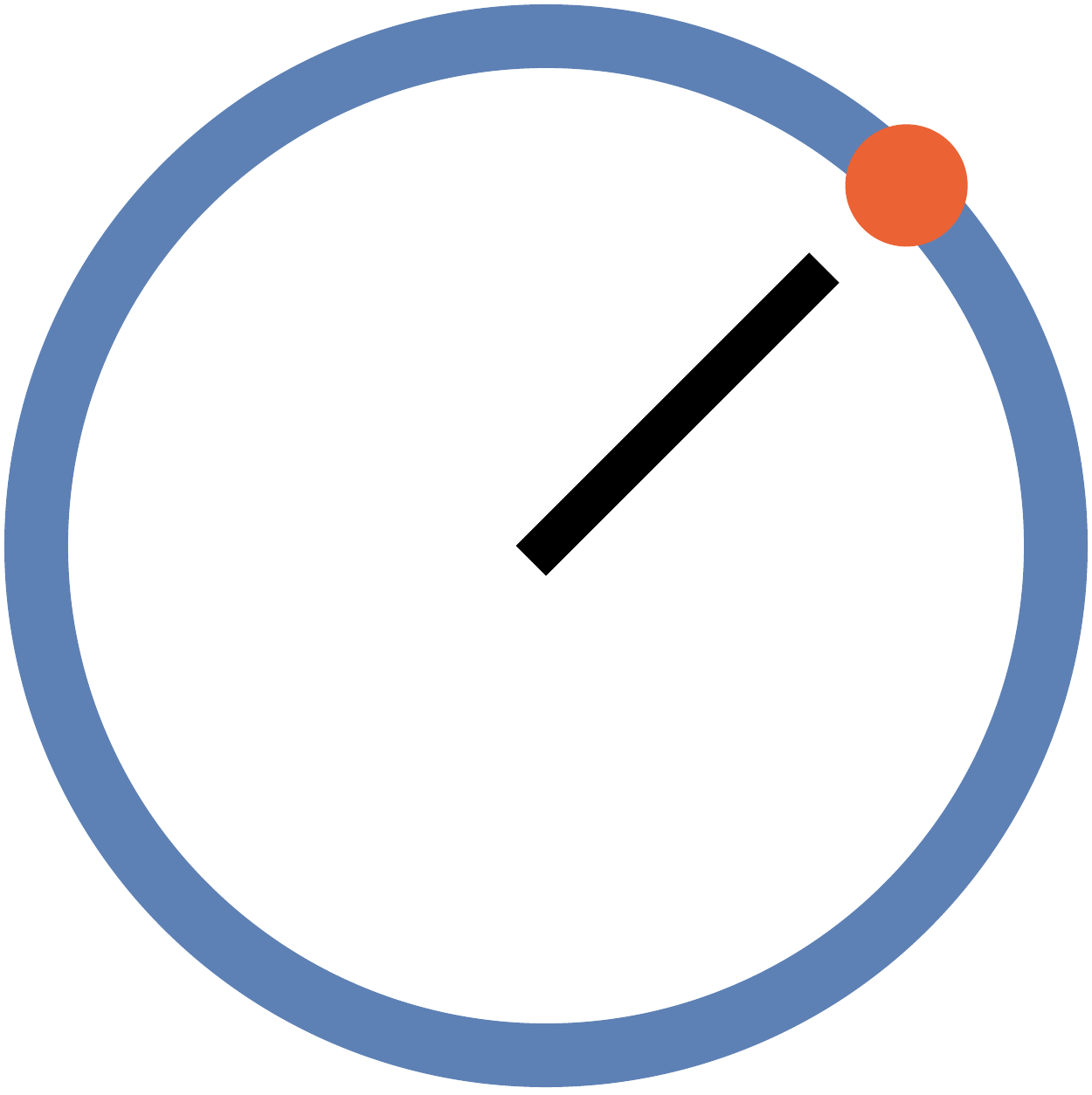}
\caption{Rotational invariance and natural metrics. A circle expresses a rotationally invariant radial distance from a central location. A natural metric can be thought of as a measure of radial distance. Different component observations that add to the same radial distance define a rotationally invariant circle.}
\label{fig:swiss}
\end{figure}

\section*{Rotational invariance}

To understand the universal scale of the normal distribution, we begin with circles and rotational invariance (\Fig{swiss}). Simple geometric concepts provide the key to natural metrics, universal scales, and the structure of commonly observed patterns.

A circle expresses a rotationally invariant radial distance from a central location. In Euclidean geometry, squared distance is the sum of squared values along each dimension. Invariant radial distance in two dimensions, $x_1$, and $x_2$, may be written as $\R^2=x_1^2+x_2^2$. The points $\left(x_1,x_2\right)$ at constant radial distance lie along the circle. The radial distance is rotationally invariant to the angle of orientation. The circular pattern is also invariant to interchange of the order of $x_1$ and $x_2$. 

We can think of the rotationally invariant circle as a way to decompose a given value into components. If we start with any observed value and equate that value with a radial distance, $\R^2$, then the observed value is equally consistent with all points $\left(x_1,x_2\right)$ that satisfy the circular constraint, $\R^2=x_1^2+x_2^2$. 

We can break up a given value into $n$ components, $\R^2=\sum x_i^2$, which is the invariant radial distance of a sphere in $n$ dimensions. Changing the order of the components does not change the radial distance. Rotational invariance implies order invariance of the component dimensions. 

\Figure{aggr} illustrates how aggregation leads to invariant distance. The top two rows aggregate randomly rotated shapes. Initially, the rows differ, because they begin with different shapes in different orientations. However, after adding many shapes, the aggregate patterns converge to the same circular form, because the order no longer matters in a large sample. The pattern of distance from the center becomes the same in every direction.

The lower two rows of \Fig{aggr} show a similar aggregate tendency to an invariant measure of distance. On the left, the initial patterns differ. As more samples are added, all information is lost except the average distance of fluctuations from the center. 

The rotational invariance of circles relates to the invariance of average distance in the normal distribution\autocite{frank16common}. In both cases, the squared distance is the standard Pythagorean definition of Euclidean geometric distance as the sum of squares. To see the connection between the rotational invariance of circles and the average distance of fluctuations in the normal distribution, we begin with an observed value and consider how it might have arisen by the aggregation of underlying components. 

\section*{Aggregation and natural metrics}

Suppose we transform an observed value, $z$, into a natural metric value, $\Tz$. What different aggregations would lead to the same value of $\Tz$? If we think of $\Tz=\Rz^2$ as a radial distance, we can evaluate the combinations of underlying values that lead invariantly to the same radial distance\autocite{frank16common}. 

Previously, we partitioned squared radial distance as
\begin{equation*}
  \Rz^2=\sum x_i^2.
\end{equation*}
We can equate the explicitly squared radial distance to the implicitly squared natural metric, $\Rz^2=\Tz$. Similarly, we can equate the explicitly squared component dimensions to the implicitly squared dimensions, $x^2 = y$, or equivalently, $x=\sqrt{y}$. Then $\Rz^2=\Tz$ can be written as
\begin{equation*}
  \Tz = \sum\sqrt{y_i}^{\,2}.
\end{equation*}
In two dimensions, the points $\left(x_1,x_2\right)$ form a circle with radius $\Rz$. The points $\pm\left(\sqrt{y}_1,\sqrt{y}_2\right)$ form an equivalent circle with radius $\Rz=\sqrt{\Tz}$.

To partition a natural metric, $\Tz$, of the observed value, $z$, we can write each component dimension, $z_i$, in its natural metric, $\T(z_i)=\T_i = y_i$, and thus
\begin{equation*}
  \Tz = \sum\sqrt{\T_i}^{\,2}.
\end{equation*}
This equation shows the different component observations of an aggregate that lead to the same rotationally invariant squared radial distance, $\Rz^2=\Tz$, or equivalently, distance as $\Rz=\sqrt{\Tz}$. 

For the natural metric, $\Tz$, the square root scale, $\sqrt{\T}$, is the natural scale of distance, aggregation, and rotational invariance.

\section*{The normal distribution}

The prior section emphasized that the natural metric $\Tz=\Rz^2$ has the square root $\sqrt{\Tz}=\Rz$ as its natural scale of distance. This section relates the normal distribution to this association between natural metrics and radial distance. See \textcite{frank16common} for additional details.

We can write the standard form of probability distributions from \Eq{expdistn} as
\begin{equation}\label{eq:normalRotate}
    \qz = ke^{-\Gl\Tz} = ke^{-\Gl\Rz^2}
\end{equation}
measured in relation to the incremental scale $\dd\sqrt{\Tz}=\dd\Rz$. Using the expression for the generalized variance, $\tilde{\Gs}^2$, in \Eq{var}, we have
\begin{equation*}
  1/2\Gl=\tilde{\Gs}^2=\ave{\T}{\sqrt{\T}}=\ave{\R^2}{\R},
\end{equation*}
and $k=\sqrt{\Gl/\pi}$. If we shift $\Tz$ so that it is expressed as a deviation from its minimum value, then for many natural metrics, $\Tz$, the probability pattern in \Eq{normalRotate} is a normal distribution with respect to the incremental scale $\dd\sqrt{\Tz}=\dd\Rz$. The distribution is centered at the minimum of $\Tz$ and has average distance of fluctuations from the central location as the generalized variance, $\tilde{\Gs}^2$. 

Different natural metrics can often be expressed in this normal form. Thus, the rotationally invariant normal form expresses a universal scale (\Fig{iversen}D).

Rotational invariance often implies invariance with respect to the order of observations in an aggregate. Order invariance connects the asymptotic rotational invariance of circles and natural metrics to the asymptotic form of the normal distribution in \Fig{aggr}. Thus, the normal distribution, expressed in natural metrics, provides a universal scale for understanding probability pattern.

\section*{Inductive: observed metric to universal scale}

How does one find natural metrics? For tree size and chemical reaction rates, I began with the observed probability pattern. From those data, I found a natural metric that fit the observed pattern. In those cases, I chose the natural metric based on the fact that patterns of size and reaction rate tend to follow a particular, commonly observed natural metric. 

This inductive approach matches a natural metric to a particular problem. The natural metric can then be used to transform the observed pattern into the universal scale of the normal distribution. What do we learn by this inductive fit of a metric and subsequent transformation to the normal form?

We have a good sense of the normal distribution as the outcome of simple aggregation and its connection to rotational invariance (\Fig{aggr}). Thus, once we find the proper scaling through the natural metric, we can think of an observed probability pattern an an expression of the normal form on a different scale. 

For example, we can think of tree size as following a normal distribution when we express size, $z$, in the natural metric $\Tz=\log(1+az)+\Gg z$. The normal form follows by expressing $\Tz$ relative to the most common size as the squared distance of a random fluctuation in relation to the distance, $\sqrt{\Tz}$. 

By recognizing the universal normal form, we can see that different measurements of the same underlying pattern express the same information. In \Fig{iversen}, the different probability patterns for rate and time have a common normal expression. Of course, many patterns that arise from unrelated processes also have the normal form. 

The key is that the structure of commonly observed pattern arises from the generic processes of aggregation and rotational invariance, when evaluated with the proper natural metric, rather than from the special attributes of particular processes. That conclusion is simply the well known principle of statistical mechanics. 

The principle of statistical mechanics is both well known and frequently ignored in the study of pattern. The reason is that the different scales on which observed patterns arise tend to obscure the underlying commonality. The point here is that one can understand natural metrics and universal scales in a rational way, and thus connect abstract principles to real problems in ways that have often been missed.

\section*{Deductive: universal scale to predicted metric}

The inductively fit metric expresses the essence of an observed pattern. But the fit does not tell us about the generative process that led to that particular metric.

Ideally, one would deduce the appropriate natural metric for a problem by considering the generative process and the necessary invariances that must be satisfied. For example, tree size must depend on growth processes, and the consequent probability pattern likely satisfies shift, stretch, and rotational invariance. However, three difficulties arise. 

First, the relations between process, measurement and pattern can be obscure. For tree size, what is the proper scale on which to measure the consequences of growth, competition, and other processes? We could use trunk diameter, $d$, or cross-sectional area, proportional to $d^2$, or a fractal exponent of diameter, $d^s$, or another size measure correlated with diameter.

The natural metric is often the scale that aggregates additively, leading to patterns that tend to be shift, stretch, and rotationally invariant. However, what we measure may be a complex transformation of that underlying scale. Inductive fit gets around the problem by describing the pattern and its associated invariant scale, rather than trying to deduce the processes that caused the observed pattern.

Second, multiple processes may shape pattern. Different processes may dominate at different scales. For example, exponential growth may dominate among smaller trees, whereas a bound on maximum size may dominate among larger trees. In general, different processes may dominate at different magnitudes. Predicting the metric that fits observations requires proper combination of the different underlying processes.

Third, natural metrics express the patterns that arise by loss of information, subject to a few minimal constraints of invariance. Because aggregation dissipates information, many seemingly distinct processes will generate the same observable pattern. Common patterns are common exactly because they match so many distinctive underlying processes \autocite{frank09the-common}. The natural metrics of common patterns reflect only the similarities of the simple invariances. Most of the special attributes of different generative processes tend to disappear in the aggregate.

\section*{Deductive: tree size example}

Tree size depends on growth, on limits to maximum size, and on a variety of other factors. Here, I give a simple introduction to natural metrics that arise from growth. I do not include bounds on size or other processes. I do not include difficulties of measurement. In spite of those limitations, this simplified analysis of growth and natural metrics provides insight into commonly observed probability patterns.

I begin with the form
\begin{equation*}
  \qz = ke^{-\Gl\Tz},
\end{equation*}
which is a normal distribution when we measure increments on the square root scale, $\dd\sqrt{\Tz}$. The normal distribution arises when we consider $\Tz$ values to be an aggregate sum of component values. 

For tree size, the problem concerns how the aggregation of random growth increments leads to the observed size. We can split total growth into $t$ increments. Each incremental unit multiplies current size by $e^{g_i}$, in which $g_i$ is the growth rate in the $i$th increment. The average growth per increment is
\begin{equation*}
  g = \frac{1}{t}\sum_{i=1}^t g_i.
\end{equation*}
Total growth is the product of all the growth increments
\begin{equation}\label{eq:ew}
  \prod e^{g_i} = e^{gt}= e^{w},
\end{equation}
in which $w=gt$ is the sum of the $t$ incremental growth rates. 

The variable $w$ provides a natural base scale for growth, because it expresses the aggregate sum of growth components. The sum is invariant to the order of the components. Thus, the total of the incremental growth rates can be thought of as a rotationally invariant radial distance.

Natural metrics arise from shift and stretch (affine) invariance to transformations of their base values \autocite{frank10measurement,frank11a-simple,frank14how-to-read}. Thus, a natural metric, $\T(w)\equiv\Tw$, for the base scale, $w$, arises from affine invariance to a generator transformation, $\G(w)$, such that 
\begin{equation*}
  \T\left[\G(w)\right] = \Ga + b\T(w)
\end{equation*}
for some constants $\Ga$ and $b$. If we consider 
\begin{equation*}
  \G(w) = \Gd + w
\end{equation*}
to be a shift of the growth rates, so that the shape of probability patterns for size does not depend on adding a constant value to growth rates, then a natural metric for size with respect to growth is
\begin{equation*}
  \Tw = e^{\Gb w},
\end{equation*}
in which $\Gb$ is a positive parameter. This metric remains affine invariant to a shift of the base scale, $w\mapsto\Gd + w$, because
\begin{equation*}
  \T\left[\G(w)\right] = e^{\Gb (\Gd+w)}= b\T(w)
\end{equation*}
for $b=e^{\Gb\Gd}$. The metric $\Tw$ is perhaps the most generic and important form of all natural metrics. Its application to growth is a special case of its underlying generality. I discussed this metric extensively in earlier articles \autocite{frank11a-simple,frank14how-to-read}. Here, I confine myself to the problem of growth in relation to size.
 
The natural metric $\Tw$ associates with the probability pattern
\begin{equation*}
  \qw = ke^{-\Gf\T_w}=ke^{-\Gf e^{\Gb w}}
\end{equation*}
when measured with respect to the incremental scale, $\dd\Tw$. If we wish to express the probability pattern with respect to measurements of growth rate, on the incremental scale $\dd w$, note that
\begin{equation*}
  \dd\Tw = \Gb\Tw\dd w = \Gb e^{\Gb w}\dd w,
\end{equation*}
yielding the probability pattern when measured with respect to the incremental base scale, $\dd w$, as
\begin{equation*}
  \qw = ke^{\Gb w -\Gf e^{\Gb w}},
\end{equation*}
in which, as always, $k$ adjusts so that the total probability is one.

Suppose we wish to transform from growth, $w$, to size, $z$, in which $w(z)$ expresses growth as a function of size. If $w$ increases with $z$, then we can write
\begin{equation*}
  \dd w = w'\dd z,
\end{equation*}
in which $w'$ is the derivative of $w$ with respect to $z$. The generic probability pattern becomes
\begin{equation}\label{eq:tzw}
  \qz = ke^{-\Gl\Tz} = ke^{\log w' + \Gb w -\Gf e^{\Gb w}}
\end{equation}
with respect to the incremental measurement scale, $\dd z$. 

In the tree size example, $w$ is the aggregate growth rate. Let $z_0+z$ be size, with $z_0$ as initial size, and $z$ as the increase in size by growth, thus
\begin{equation}\label{eq:zw}
  z_0+z = z_0 e^w,
\end{equation}
implying that $w$ as a function of $z$ is
\begin{equation}\label{eq:wz}
  w = \log(1+az).
\end{equation}
In this particular derivation, $a=1/z_0$. However, one should not interpret parameters literally. Different generative processes will lead to the same form, with alternative assumptions about process and parameters. Ultimately, the invariant properties of the metric capture the essence of common pattern. This particular derivation is meant only to show one way in which a metric arises. 

We can use \Eq{wz} to write the probability pattern of \Eq{tzw} explicitly in terms of the increase in size by growth, $z$, as
\begin{equation*}
  \qz = ke^{(\Gb-1)\log(1+az) - \Gf(1+az)^\Gb}=ke^{-\Gl\Tz}
\end{equation*}
with respect to the incremental scale, $\dd z$, yielding
\begin{equation}\label{eq:Tzdeduce}
  \Tz = \log(1+az)+\Gg(1+az)^\Gb
\end{equation}
for $\Gb<1$, and dropping constants of proportionality. For certain parameter combinations, this probability pattern will be similar to the pattern for the size metric $\Tz=\log(1+az) + \Gg z$. 

I presented this derivation to encourage future study. The proper way to relate general growth processes to invariant probability patterns remains an open problem.

\section*{Conclusion}

Probability patterns often follow a few simple scaling relations. Those scaling relations define natural metrics. A natural metric transforms measurements to a universal scale. On the universal scale, the average distance of random fluctuations from the most commonly observed value defines a generalized variance. When observed values arise by aggregation of random processes, that aggregation erases all information except the average fluctuation, the generalized variance. 

Many different probability patterns become a normal distribution when expressed on the universal scale of natural metrics. The only information in each distribution is the generalized variance. Transforming the natural metric distance back to the underlying observed values yields the standard description for probability pattern on the scale of the observed measurements. 

The great regularity of observed patterns, such as power laws, often arises from the same aspects of aggregation and invariance that lead to the normal distribution. A power law pattern and a normal distribution may simply be different transformations of the same underlying pattern. 

The transformations arise from measurement and from the invariances that define scaling relations and natural metrics \autocite{frank10measurement,frank11a-simple,frank14how-to-read,frank16common}. These key aspects of scale provide the framework in which to study the relations between pattern and process.

\subsection*{Author contributions}
SAF did all the research and wrote the article.

\subsection*{Competing interests}
No competing interests were disclosed.

\subsection*{Grant information}
National Science Foundation grant DEB--1251035 supports my research.


{\small\bibliographystyle{unsrtnat}
\bibliography{size}}

\section*{Appendix: lognormal and power law distributions}

The lognormal distribution often fits well to observed patterns. For example, many size distributions with power law attributes match reasonably well to the lognormal form. Conceptually, the lognormal distribution is easily understood by simple analogy with the normal distribution and the central limit theorem. For these reasons, the lognormal distribution is frequently used in empirical study. 

In this Appendix, I briefly introduce the lognormal distribution and its relation to power law distributions. I then explain the conceptual limitations of the lognormal distribution, and why I did not mention the lognormal distribution in main text. Finally, I show how pure power laws arise from metrics given in the text.

\subsection*{Approximate lognormal size distributions}

In the last full section of this article, I described a simple model of growth and consequent size. I split total growth into $t$ increments. Each incremental unit multiplies current size by $e^{g_i}$, in which $g_i$ is the growth rate in the $i$th increment. The average growth per increment is
\begin{equation*}
  g = \frac{1}{t}\sum_{i=1}^t g_i.
\end{equation*}
Total growth is the product of all the growth increments
\begin{equation*}
  \prod e^{g_i} = e^{gt}= e^{w},
\end{equation*}
in which $w=gt$ is the sum of the $t$ incremental growth rates. Given the total growth over the increments, we can write final size relative to initial size from \Eq{zw} as 
\begin{equation*}
  y = \frac{z_0+z}{z_0} = 1+az = e^w.
\end{equation*}
If growth in each increment, $g_i$, is a random variable, then $w$ is the sum of random growth variables. The sum of the growth variables will sometimes converge to an approximately normal distribution by the central limit theorem. The approximation to normality may be close or far off depending on particular aspects of the growth process. 

The lognormal distribution is defined as the exponential of a normal distribution. Thus, if $w$ is normally distributed, then size, $y= e^w$, has a lognormal distribution of the form
\begin{equation*}
  q_y = ke^{-\Gl(\log y - \Gm)^2-\log y}
\end{equation*}
on the incremental scale $\dd y$, in which $\Gl=1/2\Gs^2$ with $\Gs^2$ as the variance of the normal distribution in $y$. 

Lognormal distributions sometimes match reasonably well to power law patterns. As the variance in the underlying normal distribution becomes large, $\Gl$ becomes small, and the lognormal distribution becomes approximately
\begin{equation*}
  q_y \approx ke^{-\log y}=ky^{-1}.
\end{equation*}
This limiting form approximates a power law, because the log-log plot of $\log q_y$ versus $\log y$ is approximately a straight line with a slope of minus one.

\subsection*{Conceptual limitations of the lognormal}

I did not mention the lognormal distribution in the text because of its conceptual limitations. The lognormal distribution is identical to the normal distribution. If a variable, $w$, is normally distributed, then the variable $y=e^w$ has a lognormal distribution. Conceptually, there is no difference between the normal and lognormal distributions. If a pattern is normal it is also lognormal, and vice versa. Nothing is gained or lost by using one form or the other. 

In this article, I showed how one can change a wide variety of different distributions into the normal distribution. My changes truly altered the patterns of the different distributions, showing their broad conceptual unity. For example, if one has a gamma distribution
\begin{equation*}
  \qz = ke^{-\Gl(\log z + \Gg z)}=ke^{-\Gl\Tz},
\end{equation*}
then plotting $\qz$ versus $\pm\sqrt{\Tz}$ leads to a normal distribution. Here, I have related the gamma pattern to the normal form, relating two distinct probability patterns to each other. A similar approach works for many different metrics, $\Tz$.

The value in relating distributions in this way arises from two aspects. First, the common form of distributions in terms of metrics $\Tz$ follows from the two simple invariances of shift and stretch. Second, the relations between different distributions and the common normal form arises from the third invariance of rotation. Those three invariances together provide a unified framework for understanding commonly observed pattern. The lognormal does not add to that understanding, because it simply expresses the normal pattern in a slightly different way, therefore sitting outside of the conceptual framing that is the topic of this work.

\subsection*{Pure power laws}

The metrics in \Eq{Tz} and \Eq{Tzdeduce} include pure power law forms as special cases. As $\Gg\rightarrow0$, the metrics become $\Tz\rightarrow\log(1+az)$, and the distribution becomes
\begin{equation*}
  \qz = k(1+az)^{-\Gl},
\end{equation*}
which is the classic Lomax or Pareto Type II distribution. That distribution is a power law for large values of $z$. As $a$ becomes large, the distributions become a pure power law form
\begin{equation*}
  \qz=kz^{-\Gl},
\end{equation*}
in which $k$ adjusts to maintain a total probability of one.

\end{document}

%% file: size-def.tex
\newcommand*{\T}{T}
\newcommand*{\R}{R}
\newcommand*{\G}{G}

\newcommand*{\Tz}{{\T}_z}
\newcommand*{\Tr}{{\T}_r}
\newcommand*{\Tw}{{\T}_w}
\newcommand*{\Tt}{{\T}_{\Gt}}
\newcommand*{\Rz}{{\R}_z}
\newcommand*{\qz}{q_z}
\newcommand*{\qw}{q_w}
\newcommand*{\surp}{{\cal{S}}}
\newcommand*{\sz}{\surp_z}

\newcommand*{\mmskip}{\mskip1mu}

\newcommand*{\ave}[2]{\angb{#1}_{#2}}

\newcommand*{\Ga}{\alpha}
\newcommand*{\Gb}{\beta}
\newcommand*{\Gd}{\delta}

\newcommand*{\Gg}{\gamma}

\newcommand*{\Gl}{\lambda}

\newcommand*{\Gm}{\mu}

\newcommand*{\Gs}{\sigma}
\newcommand*{\Gt}{\tau}

\newcommand*{\Gf}{\phi}

\newcommand*{\angb}[1]{\left\langle#1\right\rangle}

\newcommand*{\dd}{\textrm{d}}

\newcommand*{\Eq}[1]{eqn~\ref{eq:#1}}

\newcommand*{\Figure}[1]{Figure~\ref{fig:#1}}
\newcommand*{\Fig}[1]{Fig.~\ref{fig:#1}}

\newcount\BoxNum \BoxNum 1\relax
\makeatletter
\newcommand*{\boxlabel}[1]{%
  \protected@write \@auxout {}{\string \newlabel {box:#1}{{\the\BoxNum}}{}}%
  \advance\BoxNum 1\relax}
\makeatother

\newcommand*{\boldrule}{\hrule height 1.2pt}
\newcommand*{\noterule}{\medskip\boldrule\medskip}	